\documentclass[aps,prl,reprint,groupedaddress]{revtex4-2}
\usepackage{graphicx}
\usepackage{svg}
\svgsetup{inkscapelatex=true}
\usepackage{geometry}
\usepackage{amsmath}
\usepackage{amscd}
\usepackage{amsthm}
\usepackage{amsfonts}
\usepackage{amssymb}
\usepackage{wasysym}
\usepackage{xcolor}
\definecolor{plotblue}{RGB}{0,110,191}
\definecolor{plotred}{RGB}{196,0,110}

\begin{document}

\title{Kinematic reversibility in a low Reynolds number cold atom fluid}

\author{Sara Sloman}
\author{J. Van Butcher}
\author{Chandra Raman}

\affiliation{School of Physics, Georgia Institute of Technology, 837 State St., Atlanta, GA 30332}

\date{\today}

\begin{abstract}
We have investigated kinematic reversibility in a cold atom system under strongly overdamped conditions.  In such systems, inertia is negligible, and for noninteracting rigid particles, inverting the external force causes a perfect reversal of individual particle trajectories.  We used a magneto-optical trap (MOT) as a model low Reynolds number fluid and show that kinematic reversibility survives in the presence of interparticle interactions.  In our experiment, we applied controlled external forces via a linearly ramped magnetic bias field and monitored the resulting cloud dynamics. Despite the complex three-dimensional rearrangement induced by the forces, the system exhibits precise reversibility when the force is reversed, consistent with Purcell’s framework for kinematic reversibility in low Reynolds number hydrodynamics. Reversibility was not universal, however--under certain MOT alignment conditions we have also observed clear deviations associated with system hysteresis.  Our work shows that strongly dissipative cold atom fluids are a versatile and rich platform for exploring overdamped dynamics.
\end{abstract}

\maketitle

\section{Introduction}

When a rigid object moves in an overdamped fluid under the application of an external force, its trajectory may be exactly reversed by simply inverting that force.  This is known as Kinematic Reversibility (KR).  KR is a foundational concept in the physics of fluids, and is especially applicable in the study of biological phenomena where the Reynolds number in aqueous solutions is low.  Microscopic objects, including cells, must overcome kinematic reversibility to move and, therefore, to survive in a dynamic environment \cite{RN5}.  A famous example outlined by Purcell is the 3 link swimmer \cite{RN1,RN2,RN4,RN3} where at least 2 degrees of freedom are needed to break the reversibility to achieve unidirectional motion.  Soft matter systems have allowed for the study of the transition from reversible to irreversible behavior as a function of the strength of particle interactions \cite{Corte2008RandomOI, pine2005chaos}.

Cold atom experiments have been a paradigm for the study of  controllable interactions.  However, traditional cold atom quantum fluids such as Bose and Fermi gases near zero temperature \cite{landau1987fluid} usually operate in a dissipationless regime where viscosity plays no role.  In this work we remedy this deficit by presenting the first observation of KR in a cold atom system, one where radiative forces play the role of the the fluid to create viscous damping for atomic particles \cite{PhysRevLett.59.2631, RevModPhys.71.1}.  Similar to the soft matter examples cited above, our experiments shed light on the role of interactions between particles in KR.  We create scenarios where both reversible {\em and} irreversible flows can occur depending on how the MOT is configured, thus demonstrating that KR is by no means a universal phenomena .  

A simple picture of KR as it applies to single particles in a viscous fluid can be understood as follows.  In contrast to microscopic reversibility that arises from time-reversal symmetry, KR occurs in situations where inertia can be neglected (low Reynolds number).  In this case, Newton's 2nd law, $\vec{F}_e -\gamma \vec{v}= m \frac{d\vec{v}}{dt}\approx0$, may be solved for the steady-state velocity of an object, $\vec{v}$, in the presence of  an externally applied force $\vec{F}_{e}$, yielding
\begin{equation}
    \vec{v}=\vec{F}_{e}/\gamma
\end{equation}
where $\gamma$ is the damping coefficient (the coefficient of Stokes drag).  Thus transforming $\vec{F}_e\rightarrow-\vec{F}_e$ causes $\vec{v}\rightarrow-\vec{v}$ and the particle's trajectory is exactly reversed.  Therefore, time plays no role in the behavior of the system, only its configuration.  

In spite of this simple picture, it is not at all obvious that a complex, interacting collection of particles will be subject to kinematic reversibility, since a change in sign of {\em external} forces does not imply that {\em internal} forces of interaction will reverse.  For instance, a bent piece of string pulled through honey will straighten out.  Howeveer, it retains no memory of its original shape when the force is reversed, preferring to bend in some new direction.  Cold atoms in a magneto-optical trap, however, do appear to maintain a memory as we show, due to the intrinsic balance between interactions and external forces.  Thus we have observed KR on timescales ranging from tens of milliseconds up to 10 seconds.  We also show that KR is not a universal property of interatomic forces, and a breakdown of KR concomitant with significant system hysteresis can be observed \cite{Bernstein1999,sethna2017deformation}. 
  
\begin{figure}[htbp]
\includegraphics[width= \linewidth]
{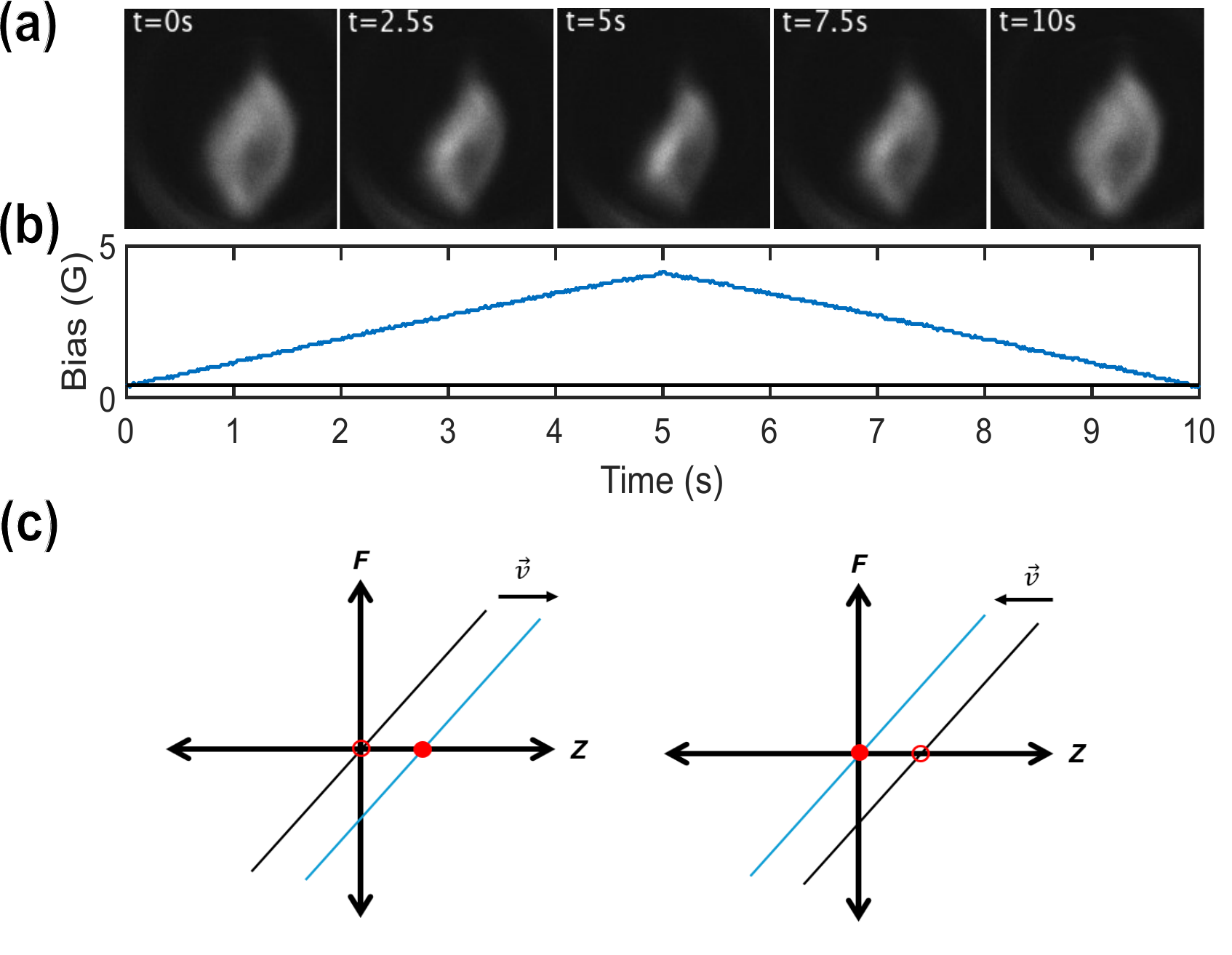}
\caption{\label{fig:magfield} \textbf{Observation of kinematic reversibility in a cold atom cloud}. (a):  A sample of stills from a video of the MOT dynamics at various times during the reversible force ramp.  (b):  Magnetic field of bias coil versus time. (c): Demonstration of reversible forces in the system.}
\end{figure}

\section{Experimental protocol}
Much work has been done to characterize the properties of magneto-optically trapped atoms (MOTs), including early models in a steady-state regime \cite{PhysRevLett.55.48,PhysRevLett.59.2631,RN7,RN9,PhysRevA.47.R4563} and instabilities and dynamical behaviors associated with the light-induced interactions between atoms \cite{stefano_verkerk_hennequin_2004, RN11, RN12, PhysRevLett.85.1839, RN10, PhysRevLett.96.023003,Haas_soares_2022,PhysRevLett.108.033001}. Here, we work with mostly stable sodium MOTs where such interactions are balanced by the external forces due to the laser beams.  Due to the strong damping caused by optical molasses \cite{PhysRevLett.55.48,PhysRevLett.59.2631}, a cloud of magneto-optically trapped atoms (a MOT) behaves as a low Reynolds number fluid, much like the biological examples cited above.  In our experiment we probe this reversibility by dragging the MOT through the viscous field of the laser beams at different rates.  Figure \ref{fig:magfield} qualitatively demonstrates our key observation of reversibility. A 6-beam MOT was created with detuning -14.8 MHz from resonance and a pair of anti-Helmholtz coils with gradient $B'=$7.7 G/cm.  We used a bias coil whose vertical axis was perpendicular  to the viewing axis of the camera to control the zero of the magnetic field, and therefore the MOT center. The location $z_0$ of the zero varies as $z_0(t)=-\frac{B_0(t)}{B'}$
where $B_0(t)$ is shown in Fig.\ \ref{fig:magfield} for $T=10$s.  

The atoms are subject to both the laser trapping force that pushes inward as well as the radiative rescattering force that pushes outward \cite{RN7,RN9,PhysRevA.47.R4563}.  The trapping force is given by the expression $F_{e}=-k(z-z_0(t))$ where $k$ is the spring constant.  Combining this with the magnetic field variation yields 
\begin{equation}
    \vec{F}_{e}=-kz - \frac{k \Delta B}{B'} \cdot\begin{cases}
        t/T & 0<t<T\\
        2-t/T&T<t<2T
    \end{cases}
\end{equation}
where $\Delta B = 3.8$ Gauss was the total change in magnetic field.  Thus, during the time $0<t<T$ the external MOT force followed the bias variation, while the radiative rescattering force dynamically adjusted to keep the MOT near equilibrium.  Due to the complex interplay of those forces that are spatially varying, the MOT underwent a 3-dimensional rearrangement as well as a translation of its center-of-mass.  Complex as it is, this rearrangement is nonetheless completely reversed 
for $T<t<2T$, demonstrating the phenomenon of kinematic reversibility.

Fluorescence images of the MOT were recorded by a camera operating at at 240 frames per second viewing from the side. The camera was triggered with the start of the magnetic field change by synchronization software. The stills in figure \ref{fig:magfield} show the distortion of the MOT between $t=0$ and $t=2T$ as the MOT is dragged by the varying magnetic field. At the maximum magnetic field (middle image), it is clear that the MOT has changed overall in size from its starting position, and that a bright spot has formed on the left side of the MOT. Comparing images from complementary times symmetric about time $T$, however, shows MOTs that appear nearly identical in size and shape. This observation is the central result of our paper, which we now quantify in detail.

\begin{figure}[!htbp]
\includegraphics[width= \linewidth]{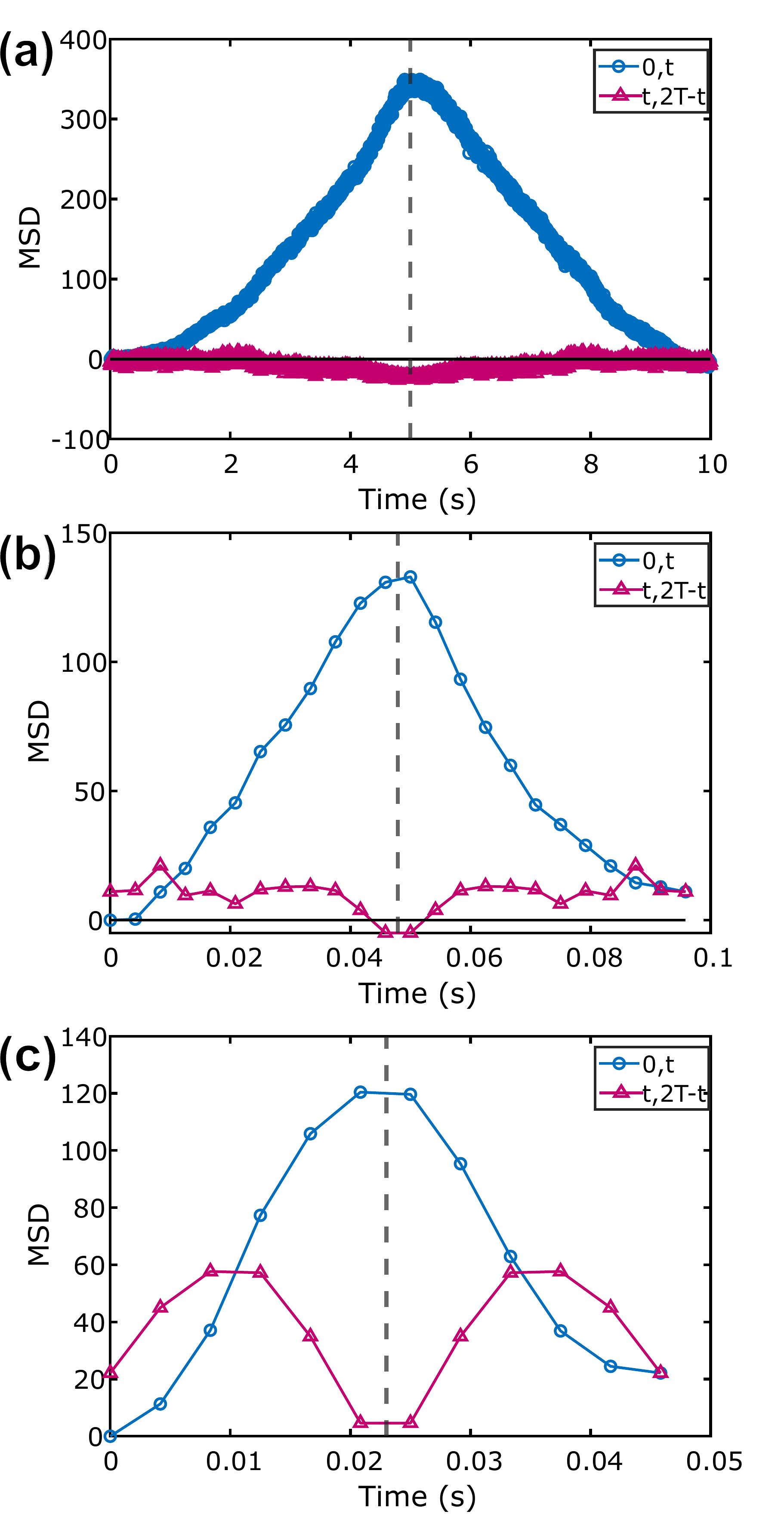}
\caption{\label{fig:msdplots} \textbf{Quantifying reversibility.} Blue circles and red triangles describe results from the (0,t) and (t,2T-t) mean-squared deviation (MSD) analysis methods, respectively. MSD is shown for MOTs moved over $2T=10$s (a), $2T=100$ms (b), and $2T=50$ms (c). Solid black line represents ideal reversibility (MSD = 0). Dashed line marks the center time.
}
\end{figure}

\section{Quantifying Reversibility}
To analyze the data we utilized the Mean-Squared Deviation (MSD), defined as 
\begin{equation}
    \textrm{MSD}(f_1,f_2)=\frac{1}{N}\sum_{x=1}^N(f_1(x)-f_2(x))^2,
\end{equation}
which compares the intensity of fluorescence collected at pixel $x$ between any two frames $f_1$ and $f_2$,  normalized to the total number of pixels $N$. .  In figure \ref{fig:msdplots}, the plotted MSD values are averaged over 10 repetitions of the experiment. Here, we use two MSD analysis methods, a $(0,t)$ method and a $(t,2T-t)$ method, which are described below. In both methods, we subtract the output of a control experiment to eliminate background fluctuations.  Further details on image processing can be found in the supplement \cite{suppMaterial}. 

The $(0,t)$  method compares the image at time $t=0$ to every subsequent image in the series at times $t>0$. Thus it quantifies the degree of change that was made.  For reference, the MSD of a MOT with no atoms compared to a fully loaded MOT was 2600, while the max MSD$(0,t)$ we observed was around 350 \cite{suppMaterial}.  In the case of perfect reversibility the MSD would start at zero at $t=0$ and end again at zero at $t=2T$, with a maximum value at $t=T$, which is what is observed in our experimental data.  Note that for perfect reversibility MSD$(0,t)=$MSD$(0,2T-t)$, which produces a curve that is symmetric about the center time. Experimental results plotted in figure \ref{fig:msdplots} of the $(0,t)$ comparison method largely follow this prediction. 

The $(t,2T-t)$ comparison quantifies the similarity of the MOT at times symmetric about time $T$. For a perfectly reversible MOT, we expect the MSD at $(t,2T-t)$ to be 0 for all $t$ since the forward and reverse trajectories would exactly align. Experimental results plotted in figure \ref{fig:msdplots} remain small when compared to the maximum distortion described by the $(0,t)$ method, demonstrating this reversibility. The maximum value of $(t,2T-t)$ decreases as $2T$ increases, meaning the MOT becomes more reversible for longer ramp times, a feature that we discuss in detail in the next section. We do note that in Fig.\ \ref{fig:msdplots}  MSD$(t,2T-t)<0$ at some times.  This is an artifact of subtracting control values, as we have explained in the supplement \cite{suppMaterial}.

\begin{figure}[htbp]
\includegraphics[width=\linewidth]{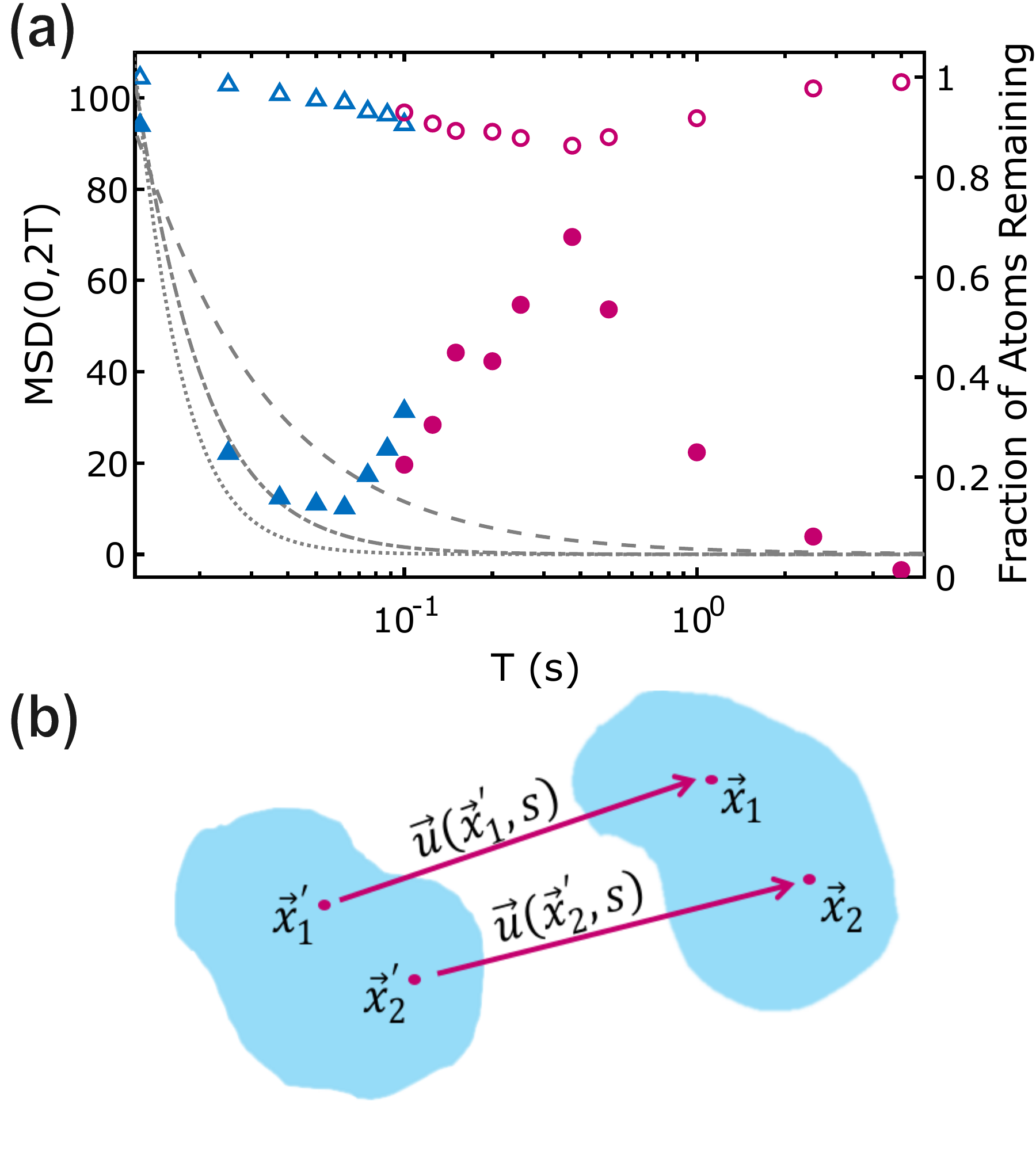}
\caption{\label{fig:timescales} \textbf{The timescales of reversibility.} (a): Blue triangles (\textcolor{plotblue}{$\blacktriangle$} ,\textcolor{plotblue}{$\triangle$}) are data that  taken when loading was turned off. Red circles (\textcolor{plotred}{$\CIRCLE$},\textcolor{plotred}{$\Circle$}) are data   taken with loading turned on. Filled shapes show the average $MSD(0,2T)$ for each time period at which the data was taken (left axis) and outlined shapes describe the fractional change in number of atoms from start to finish versus time (right axis). Dashed curves are trends proportional to $1/T$ (large dashes), $1/T^2$ (mixed size dashes) and $1/T^3$ (small dashes). (b): Schematic of the deformation which occurs due to the finite time response. Red arrows represent the displacement vector $\vec{u}$ that maps initial ($\vec{x}'_1,\vec{x}'_2$) to final ($\vec{x}_1,\vec{x}_2$) positions of a particle in the MOT that has undergone distortion proportional to parameter $s$. For perfect reversibility $u\rightarrow 0$. 
}
\end{figure}

\section{MODELING THE APPROACH TO REVERSIBILITY}

In our system, reversibility is dependent on the atomic density distribution being close to steady state at all times.  Therefore if the forces are applied too quickly the MSD will increase due to the finite time response of the cloud.  In Fig.\ \ref{fig:timescales} we have quantified this behavior.  We varied $T$ from 25ms to 10 seconds in a series of experiments, computing MSD$(0,2T)$ at each point. In order to distinguish the influence of MOT loss and reloading dynamics, the loading of the MOT was shut off for $T\leq 100$ ms, while for longer times it was left on to counter the effect of atom loss.  Our data is therefore a composite of the two regimes that we must differentiate in our analysis.

The overall trend in Fig.\ \ref{fig:timescales} is a decrease in MSD with time starting from 25 ms.  This trend is complicated by the loss/loading dynamics mentioned, causing a `bump' where the MSD begins to increase around 0.07 s due to accumulating loss, and then decreases for times $> 0.4$ s as the loading compensates the loss.  This is also reflected in the data showing the fraction of remaining atoms, which experiences a clear dip during the time periods where the MSD `bump' occurs.  At longer times, the MSD decreases to near zero.  Apart from this loading-related bump that is unavoidable for our system, the data as a whole appears to follow a steadily decreasing MSD trend proportional to $T^{-2}$ (for comparison, both $T^{-1}$ and $T^{-3}$ are shown, both of which fit the data less well than $T^{-2}$). These trend lines are based on the microscopic model described in the supplement \cite{suppMaterial}, and have been scaled to fit the data. We turn now to an explanation of this trend.  

For perfect reversibility, in the absence of any loss and absent inertial effects, the overdamped atomic fluid should return exactly to its initial state.  We model deviations from perfect reversibility due to the finite time response by a deformation parameter $s$ for the density profile $n(\vec{x},s)$ that is measured in the experiment through fluorescence imaging.  Since a very slow ramp reduces the deviation between initial and final states, we expect that $s\rightarrow 0$ as $T\rightarrow\infty$.  
The deformation is related to a nonzero value for the fluid displacement field $\vec{u}(\vec{x}',s)$ that is shown in Fig.\ \ref{fig:timescales}b \cite{landau1987fluid}.  $\vec{u}$ is the difference between initial $\vec{x}'$ and final $\vec{x}$ positions of a particle in the fluid after the ramp, which can be expanded in the small parameter $s$:
\begin{equation}
    \vec{u}(\vec{x'},s)\equiv\vec{x}-\vec{x}' = s \vec{u}_1(\vec{x}')+O(s^2)
    \label{eq:displacement_field}
\end{equation}
Note that for perfect reversibility $\vec{u}=0$ as every particle will have returned to its original position.  The coefficient of $s$ on the rightmost side of Eqn.\ \ref{eq:displacement_field} is a ``velocity'' field, $\vec{u}_1 \equiv [\partial\vec{u}/\partial s]_{s=0}$, the infinitesimal displacement per unit $s$, evaluated at zero deformation.  Thus $\vec{u}_1$ is only a property of the asymptotic distribution $n(\vec{x},s\rightarrow 0$).  The full velocity field $\vec{v}(\vec{x},s)  =\partial \vec{u}/\partial s$ and fluid density $n(\vec{x},s)$ obey the continuity equation
\begin{equation}
    \frac{\partial n}{\partial s} + \nabla \cdot \big( n\,\vec{v} \big) =0
    \label{eq:continuity}
\end{equation}
Expanding $n(\vec{x},s)=n(\vec{x},0)+s\left(\partial n/\partial s\right)_{s=0} +O(s^2)$ for small deformations and setting $\vec{v}=\vec{u}_1$, we obtain, using Eqn.\ \ref{eq:continuity},
\begin{equation}
n(\vec{x},s)
=
n(\vec{x},0)
-
s\,\nabla \cdot\big( n(\vec{x},0)\,\vec{u}_1(\vec{x})\big)
+
O(s^2).
\label{eq:continuity-pred}
\end{equation}
Provided there is little to no loss of atoms during the ramp, we can compute the MSD for small displacements by taking $n(\vec{x},0)$ to be equal to the initial density profile in the experiment, $n_0(\vec{x})$, and $n(\vec{x},s)$ to be the final one.  The difference is what enters the experimentally determined MSD:
\begin{equation}
    \textrm{MSD} = \int (n(\vec{x},s)-n_0(\vec{x}))^2 dx \approx s^2 \int q^2(x)dx
\end{equation}
to leading order in $s$, where the function $q(x)=-\nabla \cdot\big( n_0(x)\,\vec{u}_1(x)\big)$ is independent of the deformation.  In the Supplement \cite{suppMaterial} we present a microscopic model for the  displacement $u$ of a particle in the MOT due to the time-varying external force.  This model predicts that $u \rightarrow 0$ as $1/T$, and since $u\sim s$ from Eqn.\ \ref{eq:displacement_field}, the MSD decays as $s^2 \sim 1/T^2$, in good agreement with the data shown in Fig.\ \ref{fig:timescales} when we account for loading related changes to the MOT. 

While the model describes the asymptotic behavior, it does not account for the observed MSD$(0,T)$.  In particular, we observed that MSD$(0,T)$ grows continuously with $T$ (see Fig. 3 of \cite{suppMaterial}).  In contrast, the model predicts a saturation behavior.  However, the model only includes single particle dynamics and not the full range of interactions between particles, which is the likely reason for the discrepancy.  These interactions cause the cloud's motion to lag behind the applied forces.  In the next section we present a variation on the experiment that caused this lag behavior to be extremely pronounced, and the reversibility to be highly suppressed in comparison to the data shown thus far.

\begin{figure}[!htbp]
\includegraphics[width= \linewidth]{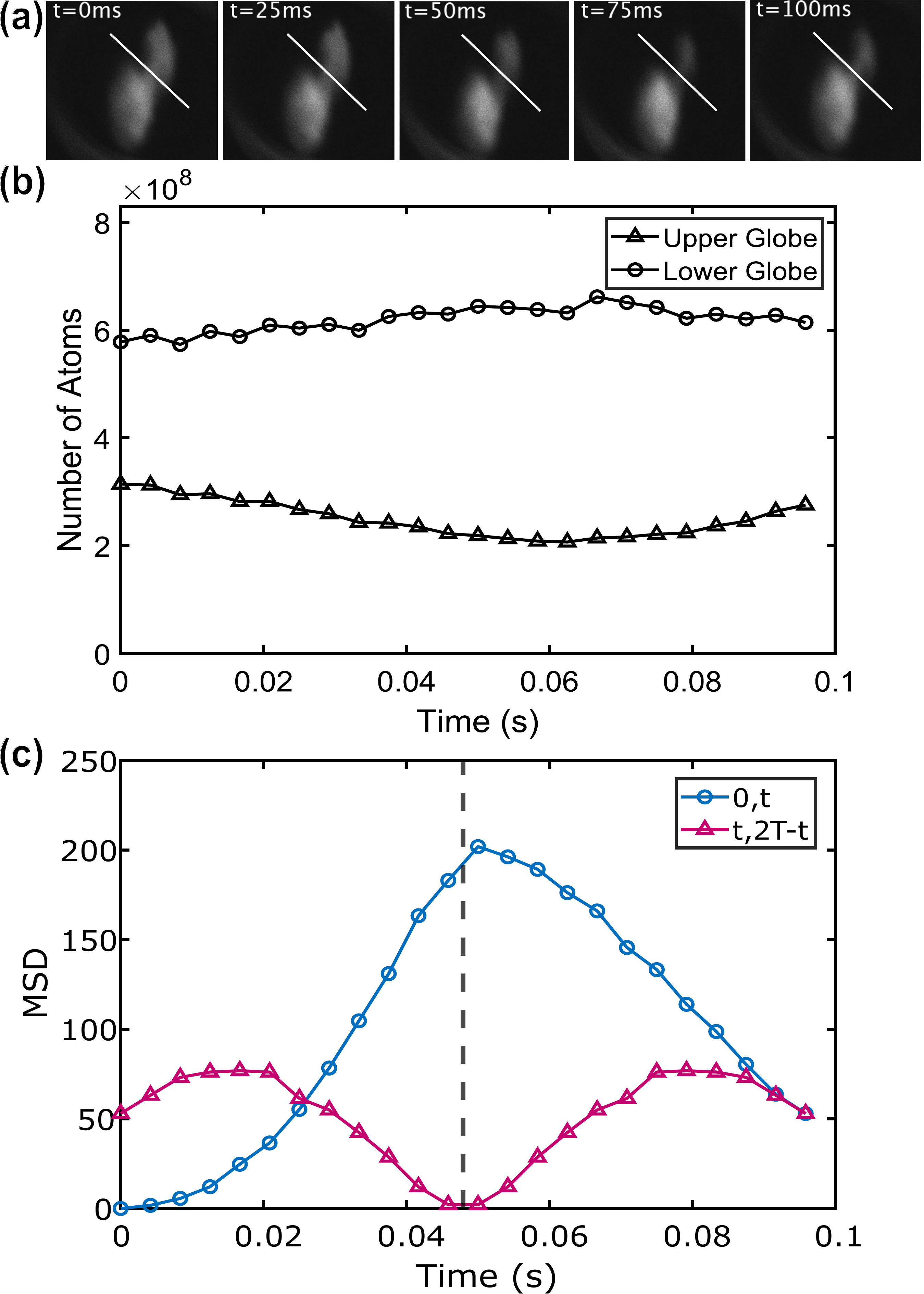}
\caption{\label{fig:hystereticmot} \textbf{Effects of Hysteresis on Reversibility.} (a): Stills from video recording of a hysteretic MOT during reversible force ramp over $2T=100$ms. (b): Comparison of atom number between two circular regions of the MOT, above and below white line. (c): The $(0,t)$ analysis method (blue) and $(t,2T-t)$ analysis method (red) applied to the entire hysteretic MOT. The
dashed black line marks the center time. }
\end{figure}

\section{Hysteretic MOT}

In Fig.\ \ref{fig:hystereticmot},  we show data from a MOT that experienced significant hysteresis when performing the same reversibility experiment. The result is a MOT that ultimately still reverses, but on a longer timescale and with unexpected behavior throughout the experimental sequence.

Typically, well aligned laser beams will balance the forces in a MOT, creating a spherical cloud of atoms. Poorly aligned laser beams will prevent a MOT from forming, but if the misalignment is slight, a MOT can still form, but is often misshapen due to the imbalance of forces. Fig.\ \ref{fig:hystereticmot}a is one such example of these slightly misaligned MOTs, where the cloud has divided into two distinct yet connected globes, one above the other in the image. The white line was placed in the images to divide upper and lower globes.  The reversibility experiment then caused atoms to be first transferred from upper to lower globes in the time interval $0<t<50$ms, and transferred partially back for $50<t<100$ms.  A qualitative comparison of Fig.\ \ref{fig:magfield}a and Fig.\ \ref{fig:hystereticmot}a demonstrates that the hysteric MOT does not fully regain its initial form throughout the experiment, while the MOTs previously described appear to return entirely to their initial position. 

The data in Fig.\ \ref{fig:hystereticmot}b shows atoms are moving from the top globe to the bottom globe as they are dragged vertically, but the number of atoms in each globe does not return to its original value. A closer look reveals that this behavior cannot be due to a global loss of atoms.  In fact, in Fig.\ \ref{fig:hystereticmot}b, we can see that while the upper globe experiences an overall loss of atoms, as is expected in this experiment when loading is not present, the lower globe experiences a net gain of atoms. Here, atom number is estimated based on the fluorescence and scattering rate of the atoms.

For this MOT, the MSD$(0,t)$ shown in Fig.\ \ref{fig:hystereticmot}c for $2T=100$ms does not yield a curve that is symmetric about the center time. By comparison, the data in Fig.\ \ref{fig:msdplots}b at $2T=100$ms was also taken without loading but shows a much higher degree of symmetry about the center. Fig.\ \ref{fig:timescales} has MSD$(0,2T=100)\approx 11$ whereas in Fig.\ \ref{fig:hystereticmot} it can be seen to be larger, $\approx 50$.  The maximum value of MSD$(t,2T-t)$ was 76.8, which is a factor of 3.6 greater than the prior data, indicating much lower reversibility.  Data taken for larger values of $T$ showed that the system was returning to the initial state, albeit more slowly.

The data in Fig.\ \ref{fig:hystereticmot}c show a clear hysteresis in the system behavior, where the timescale for reverse motion becomes considerably longer than that of forward motion.  Fig.\ \ref{fig:hystereticmot}c shows that the rate of change of MSD$(0,t)$ is much lower in the 2nd half of the scan compared with the first half. This appears to be a form of jamming behavior \cite{keys2007measurement}, where the MOT resists coming back to the upper globe due to a reduced mobility. We estimate the total number of atoms in the hysteretic MOT to be about $8.8\times10^9$ atoms (for details, see \cite{suppMaterial}), which is well above the threshold at which the atom density has become capped by rescattering forces \cite{PhysRevLett.70.2253}.  Thus the forward and reverse flows are that of an incompressible {\em atomic} fluid.  Any attempt to reverse the external force that requires a further increase in density would be opposed by the fluid.  This could be a mechanism for the observed jamming behavior. This implies that irreversibility may be linked to the peculiar MOT geometry, specifically the balance of forces that causes the splitting into two separated globes.  Further exploration of this phenomenon is warranted.

\section{Discussion}
We have experimentally demonstrated kinematic reversibility (KR) in a cold atom system.  We have also provided a dynamical model that agrees well with our experimental data and identifies key scaling behavior of the system's approach to reversibility.  Our work shows that when a MOT is well aligned and stable, it exhibits reversibility, but that reversibility is by no means a universal feature of this cold atom system. In particular, the observation of MOT hysteresis opens up possibilities for future experiments that modify the geometry, e.g. using spatial light modulators, to induce and control jamming, and to explore phase transitions between reversible and irreversible behavior.

\section{Acknowledgments}
This work was supported by National Science Foundation Award no. 2011478. We thank Aniruddha Bhattacharya for his discussions at the early stages of this project.

\bibliography{PaperDraft}

@article{RN2,
   author = {Becker, L. E. and Koehler, S. A. and Stone, H. A.},
   title = {On self-propulsion of micro-machines at low Reynolds number: Purcells three-link swimmer},
   journal = {Journal of Fluid Mechanics},
   volume = {490},
   pages = {15-35},
   ISSN = {00221120
14697645},
   DOI = {10.1017/s0022112003005184},
   year = {2003},
   type = {Journal Article}
}

@article{RN3,
   author = {Gagnon, D. A. and Arratia, P. E.},
   title = {The cost of swimming in generalized Newtonian fluids: experiments withC.elegans},
   journal = {Journal of Fluid Mechanics},
   volume = {800},
   pages = {753-765},
   ISSN = {0022-1120
1469-7645},
   DOI = {10.1017/jfm.2016.420},
   url = {https://www.cambridge.org/core/services/aop-cambridge-core/content/view/D83E96465B27C09731505D629551A7AE/S0022112016004201a.pdf/div-class-title-the-cost-of-swimming-in-generalized-newtonian-fluids-experiments-with-span-class-italic-c-elegans-span-div.pdf},
   year = {2016},
   type = {Journal Article}
}

@article{RN10,
   author = {Gaudesius, M. and Kaiser, R. and Labeyrie, G. and Zhang, Y. C. and Pohl, T.},
   title = {Instability threshold in a large balanced magneto-optical trap},
   journal = {Physical Review A},
   volume = {101},
   number = {5},
   ISSN = {2469-9926},
   DOI = {ARTN 053626
10.1103/PhysRevA.101.053626},
   url = {<Go to ISI>://WOS:000533490700015},
   year = {2020},
   type = {Journal Article}
}

@article{RN11,
   author = {Gaudesius, M. and Zhang, Y. C. and Pohl, T. and Kaiser, R. and Labeyrie, G.},
   title = {Phase diagram of spatiotemporal instabilities in a large magneto-optical trap},
   journal = {Physical Review A},
   volume = {103},
   number = {4},
   ISSN = {2469-9926
2469-9934},
   DOI = {10.1103/PhysRevA.103.L041101},
   url = {https://journals.aps.org/pra/pdf/10.1103/PhysRevA.103.L041101},
   year = {2021},
   type = {Journal Article}
}

@article{RN12,
   author = {Gaudesius, M. and Zhang, Y. C. and Pohl, T. and Kaiser, R. and Labeyrie, G.},
   title = {Three-dimensional simulations of spatiotemporal instabilities in a magneto-optical trap},
   journal = {Physical Review A},
   volume = {105},
   number = {1},
   ISSN = {2469-9926
2469-9934},
   DOI = {10.1103/PhysRevA.105.013112},
   url = {https://journals.aps.org/pra/pdf/10.1103/PhysRevA.105.013112},
   year = {2022},
   type = {Journal Article}
}

@article{RN4,
   author = {Hubert, M. and Trosman, O. and Collard, Y. and Sukhov, A. and Harting, J. and Vandewalle, N. and Smith, A. S.},
   title = {Scallop Theorem and Swimming at the Mesoscale},
   journal = {Phys Rev Lett},
   volume = {126},
   number = {22},
   pages = {224501},
   ISSN = {1079-7114 (Electronic)
0031-9007 (Linking)},
   DOI = {10.1103/PhysRevLett.126.224501},
   url = {https://www.ncbi.nlm.nih.gov/pubmed/34152187
https://orbi.uliege.be/bitstream/2268/260700/1/PhysRevLett.126.224501.pdf},
   year = {2021},
   type = {Journal Article}
}

@article{RN5,
   author = {Lauga, Eric and Powers, Thomas R.},
   title = {The hydrodynamics of swimming microorganisms},
   journal = {Reports on Progress in Physics},
   volume = {72},
   number = {9},
   ISSN = {0034-4885
1361-6633},
   DOI = {10.1088/0034-4885/72/9/096601},
   year = {2009},
   type = {Journal Article}
}

@article{RN1,
   author = {Purcell, E. M.},
   title = {Life at low Reynolds number},
   journal = {American Journal of Physics},
   volume = {45},
   number = {1},
   pages = {3-11},
   ISSN = {0002-9505
1943-2909},
   DOI = {10.1119/1.10903},
year={1977},
   type = {Journal Article}
}

@article{RN9,
   author = {Sesko, D. W. and Walker, T. G. and Wieman, C. E.},
   title = {Behavior of neutral atoms in a spontaneous force trap},
   journal = {Journal of the Optical Society of America B},
   volume = {8},
   number = {5},
   ISSN = {0740-3224
1520-8540},
   DOI = {10.1364/josab.8.000946},
   year = {1991},
   type = {Journal Article}
}

@article{RN7,
   author = {Walker, T. and Sesko, D. and Wieman, C.},
   title = {Collective behavior of optically trapped neutral atoms},
   journal = {Phys Rev Lett},
   volume = {64},
   number = {4},
   pages = {408-411},

   ISSN = {1079-7114 (Electronic)
0031-9007 (Linking)},
   DOI = {10.1103/PhysRevLett.64.408},
   url = {https://www.ncbi.nlm.nih.gov/pubmed/10041972
https://journals.aps.org/prl/pdf/10.1103/PhysRevLett.64.408},
   year = {1990},
   type = {Journal Article}
}

@article{PhysRevLett.59.2631,
  title = {Trapping of Neutral Sodium Atoms with Radiation Pressure},
  author = {Raab, E. L. and Prentiss, M. and Cable, Alex and Chu, Steven and Pritchard, D. E.},
  journal = {Phys. Rev. Lett.},
  volume = {59},
  issue = {23},
  pages = {2631--2634},
  numpages = {0},
  year = {1987},
  month = {Dec},
  publisher = {American Physical Society},
  doi = {10.1103/PhysRevLett.59.2631},
  url = {https://link.aps.org/doi/10.1103/PhysRevLett.59.2631}
}

@article{PhysRevLett.55.48,
  title = {Three-dimensional viscous confinement and cooling of atoms by resonance radiation pressure},
  author = {Chu, Steven and Hollberg, L. and Bjorkholm, J. E. and Cable, Alex and Ashkin, A.},
  journal = {Phys. Rev. Lett.},
  volume = {55},
  issue = {1},
  pages = {48--51},
  numpages = {0},
  year = {1985},
  month = {Jul},
  publisher = {American Physical Society},
  doi = {10.1103/PhysRevLett.55.48},
  url = {https://link.aps.org/doi/10.1103/PhysRevLett.55.48}
}

@article{PhysRevLett.70.2253,
  title = {High densities of cold atoms in a dark spontaneous-force optical trap},
  author = {Ketterle, Wolfgang and Davis, Kendall B. and Joffe, Michael A. and Martin, Alex and Pritchard, David E.},
  journal = {Phys. Rev. Lett.},
  volume = {70},
  issue = {15},
  pages = {2253--2256},
  numpages = {0},
  year = {1993},
  month = {Apr},
  publisher = {American Physical Society},
  doi = {10.1103/PhysRevLett.70.2253},
  url = {https://link.aps.org/doi/10.1103/PhysRevLett.70.2253}
}

@article{RevModPhys.71.1,
  title = {Experiments and theory in cold and ultracold collisions},
  author = {Weiner, John and Bagnato, Vanderlei S. and Zilio, Sergio and Julienne, Paul S.},
  journal = {Rev. Mod. Phys.},
  volume = {71},
  issue = {1},
  pages = {1--85},
  numpages = {0},
  year = {1999},
  month = {Jan},
  publisher = {American Physical Society},
  doi = {10.1103/RevModPhys.71.1},
  url = {https://link.aps.org/doi/10.1103/RevModPhys.71.1}
}

@article{PhysRevA.47.R4563,
  title = {Collisional loss rate in a magneto-optical trap for sodium atoms: Light-intensity dependence},
  author = {Marcassa, L. and Bagnato, V. and Wang, Y. and Tsao, C. and Weiner, J. and Dulieu, O. and Band, Y. B. and Julienne, P. S.},
  journal = {Phys. Rev. A},
  volume = {47},
  issue = {6},
  pages = {R4563--R4566},
  numpages = {0},
  year = {1993},
  month = {Jun},
  publisher = {American Physical Society},
  doi = {10.1103/PhysRevA.47.R4563},
  url = {https://link.aps.org/doi/10.1103/PhysRevA.47.R4563}
}

@article{PhysRevLett.85.1839,
  title = {Instabilities in a Magneto-optical Trap: Noise-Induced Dynamics in an Atomic System},
  author = {Wilkowski, David and Ringot, Jean and Hennequin, Daniel and Garreau, Jean Claude},
  journal = {Phys. Rev. Lett.},
  volume = {85},
  issue = {9},
  pages = {1839--1842},
  numpages = {0},
  year = {2000},
  month = {Aug},
  publisher = {American Physical Society},
  doi = {10.1103/PhysRevLett.85.1839},
  url = {https://link.aps.org/doi/10.1103/PhysRevLett.85.1839}
}

@article{PhysRevLett.96.023003,
  title = {Self-Sustained Oscillations in a Large Magneto-Optical Trap},
  author = {Labeyrie, G. and Michaud, F. and Kaiser, R.},
  journal = {Phys. Rev. Lett.},
  volume = {96},
  issue = {2},
  pages = {023003},
  numpages = {4},
  year = {2006},
  month = {Jan},
  publisher = {American Physical Society},
  doi = {10.1103/PhysRevLett.96.023003},
  url = {https://link.aps.org/doi/10.1103/PhysRevLett.96.023003}
}

@article{stefano_verkerk_hennequin_2004, 
 title={Deterministic instabilities in the magneto-optical trap}, volume={30}, 
 url={https://link.springer.com/article/10.1140/epjd/e2004-00089-y#citeas}, 
 DOI={https://doi.org/10.1140/epjd/e2004-00089-y}, 
 number={2}, 
 journal={The European Physical Journal D}, 
 publisher={Springer Science and Business Media LLC}, author={Stefano, A. and Verkerk, Ph. and Hennequin, D.}, year={2004}, 
 month={Aug}, 
 pages={243–258} }

@book{landau1987fluid,
  title={Fluid Mechanics},
  author={Landau, L. D. and Lifshitz, E. M.},
  volume={6},
  series={Course of Theoretical Physics},
  year={1987},
  publisher={Pergamon Press},
  address={Oxford},
  edition={2nd English},
  note={Translated by J.B. Sykes and W.H. Reid}
}

@article{PhysRevLett.108.033001,
  title = {Photon Bubbles in Ultracold Matter},
  author = {Mendon\ifmmode \mbox{\c{c}}\else \c{c}\fi{}a, J. T. and Kaiser, R.},
  journal = {Phys. Rev. Lett.},
  volume = {108},
  issue = {3},
  pages = {033001},
  numpages = {5},
  year = {2012},
  month = {Jan},
  publisher = {American Physical Society},
  doi = {10.1103/PhysRevLett.108.033001},
  url = {https://link.aps.org/doi/10.1103/PhysRevLett.108.033001}
}

@Article{Haas_soares_2022,
  AUTHOR = {Haas, Fernando and Soares, Luiz Gustavo Ferreira},
  TITLE = {Nonlinear Dynamics in Isotropic and Anisotropic Magneto-Optical Traps},
  JOURNAL = {Atoms},
  VOLUME = {10},
  YEAR = {2022},
  NUMBER = {3},
  ARTICLE-NUMBER = {83},
  URL = {https://www.mdpi.com/2218-2004/10/3/83},
  DOI = {10.3390/atoms10030083},
  ISSN = {2218-2004}
}

@article{Bernstein1999,
  author = {B. Bernstein and T. Erber},
  title = {Reversibility, irreversibility: restorability, non-restorability},
  journal = {Journal of Physics A: Mathematical and General},
  volume = {32},
  number = {43},
  pages = {7581--7602},
  year = {1999},
  publisher = {IOP Publishing},
  doi = {10.1088/0305-4470/32/43/310},
  url = {https://iopscience.iop.org/article/10.1088/0305-4470/32/43/310}
}

@misc{suppMaterial,
  note = {See Supplemental Material at \url{http://link.aps.org} for details of data collection and processing, experimental resolution limits, and particle position modeling, which also references \cite{dalibard1988laser}.}
}

@article{keys2007measurement,
  title={Measurement of growing dynamical length scales and prediction of the jamming transition in a granular material},
  author={Keys, Aaron S and Abate, Adam R and Glotzer, Sharon C and Durian, Douglas J},
  journal={Nature Physics},
  volume={3},
  number={4},
  pages={260--264},
  year={2007},
  publisher={Nature Publishing Group},
  doi={10.1038/nphys572}
}

@article{pine2005chaos,
  title={Chaos and threshold for irreversibility in sheared suspensions},
  author={Pine, David J and Gollub, Jerry P and Brady, John F and Leshansky, Alexander M},
  journal={Nature},
  volume={438},
  number={7070},
  pages={997--1000},
  year={2005},
  publisher={Nature Publishing Group},
  doi={10.1038/nature04380}
}

@article{Corte2008RandomOI,
  title={Random organization in periodically driven systems},
  author={Laurent Cort{\'e} and Paul M. Chaikin and Jerry P. Gollub and David J. Pine},
  journal={Nature Physics},
  year={2008},
  volume={4},
  pages={420--424},
  doi={10.1038/nphys891}
}

@article{sethna2017deformation,
  title={Deformation of crystals: Connections with statistical physics},
  author={Sethna, James P and Bierbaum, Matthew and Dahmen, Karin A and Goodrich, Carl P and Greer, Julia R and Hayden, Lorien X and Kent-Dobias, Jaron and Lee, Edward D and Liarte, Danilo B and Ni, Xiaoyue and others},
  journal={Annual Review of Materials Research},
  volume={47},
  pages={217--246},
  year={2017},
  publisher={Annual Reviews},
  doi={10.1146/annurev-matsci-070115-032036}
}

\end{document}